\documentclass[review,5p,times,twocolumn,preprint]{elsarticle}
\usepackage{graphicx}
\usepackage{amssymb}
\usepackage{lineno}
\biboptions{numbers,sort&compress}

\journal{Thin Solid Films}

\begin{document}

\begin{frontmatter}

\title{X-ray scattering characterization of iron oxide nanoparticles Langmuir film on water surface and on a solid substrate}

\author[label1,label5]{V. Ukleev}
\address[label1]{National Research Centre "Kurchatov Institute" B. P. Konstantinov Petersburg Nuclear Physics Institute, 188300 Gatchina, Russia}
\ead{ukleev@lns.pnpi.spb.ru}
\fntext[label5]{Present address: RIKEN Center for Emergent Matter Science (CEMS), Wako 351-0198, Japan}

\author[label2]{A. Khassanov}
\address[label2]{Institute of Polymer Materials of the Department of Materials Science Friedrich-Alexander University Erlangen-N\"{u}rnberg, Martensstrasse 7, D-91058 Erlangen, Germany}
\author[label3]{I. Snigireva}
\address[label3]{European Synchrotron Radiation Facility, 71, Avenue des Martyrs, CS40220, F-38043 Grenoble Cedex 9}
\author[label3]{O. Konovalov}
\author[label3,label4]{A. Vorobiev}
\ead{avorobiev@ill.fr}
\address[label4]{Department of Physics and Astronomy, Uppsala University, Box 516, 751 20, Uppsala, Sweden}

\begin{abstract}
In the present study we compare a structure of a Langmuir film assembled from magnetic iron oxide nanoparticles on water surface and a structure of the same film after its transfer to a solid substrate by the Langmuir-Schaefer method. In contrast to most of related studies, where different techniques are used to characterize the films before and after the deposition, we use the same combination of X-ray reflectometry and Grazing Incidence Small-Angle X-ray scattering. In both cases -- on a liquid and on a solid substrate -- the film was identified as a well-ordered monolayer of the nanoparticles laterally organized in a two-dimensional hexagonal lattice. However parameters of the lattice were found to be slightly different depending on the type of the substrate. It is also demonstrated that Langmuir-Schaefer technique is the right way for deposition of such kind of the particles on a solid substrate.
\end{abstract}


\end{frontmatter}

\section{Introduction}

Ordered arrays of magnetic nanoparticles (MNPs) are promising objects for various applications in biomedicine \cite{ito2005medical, gao2009multifunctional}, catalyst \cite{rybczynski2003large, hu2005magnetically, roy2009functionalized, schatz2010nanoparticles}, optics \cite{nie2010properties} and high-density data storage \cite{terris2005nanofabricated, nie2010properties}. Furthermore, self-assembly of nanosized objects is interesting from a point of view of a fundamental understanding of the interplay between competing driving forces at the nanoscale.

Iron oxide MNPs can be assembled into a monolayer by various methods, such as drop casting \cite{sun2000monodisperse,siffalovic2007self,siffalovic2008real}, doctor blade casting \cite{bodnarchuk2009large, yang2010large} and spin coating \cite{johnston2011formation, mishra2012self}. Self-assembly of the nanoparticles occurs during the solvent evaporation, therefore interaction between the film and a substrate plays an important role in eventual ordering. It was recently shown, that ultimately large-area monolayers of iron oxide MNPs can be produced on water surface using Langmuir technique \cite{guo2003patterned, lee2007vast, pauly2009large, wen2011ultra, vorobiev2015substantial}. However, assembling of the monolayer  on liquid surface is only first technological step. For further applications an array of MNPs should be transferred from the Langmuir trough to a solid substrate what is usually done either by Langmuir-Blodgett (LB) or by Langmuir-Schaefer (LS) techniques. X-ray reflectometry (XRR) and Grazing Incidence Small-Angle X-ray scattering (GISAXS) are the most appropriate experimental methods to study the nanostructure of the Langmuir films in-situ, i.e. as they form directly on the liquid surface. While Scanning electron microscopy (SEM) is conventional way of characterization of the resulting LB- or LS-assembled nanoparticle films \cite{guo2003patterned, pauly2009large, wen2011ultra, dochter2013multiscale, meijer2012self}. It was recently observed that a lattice constant of the same nanoparticle film obtained by GISAXS on a liquid surface and by SEM after a deposition on a solid substrate can be significantly different. The difference can be caused by several factors, including imbalance of repulsive and attractive forces \cite{vorobiev2015substantial}, the transfer method \cite{aleksandrovic2008preparation} and physical properties of the particles \cite{banerjee2011structural}. Furthermore, an average structural domain size as obtained by GISAXS and SEM can have different values due to the difference in resolution and due to different areas of a sample probed by these techniques \cite{stanley2015novel}.

Therefore we used the same combination of XRR and GISAXS  to perform analysis of MNPs organization in-plane and out-of-plane of the sample  both in-situ on water subphase and ex-situ after the deposition on a solid substrate.

\section{Samples}

Iron oxide Fe$_2$O$_3$ maghemite nanoparticles with the size of 10 nm (denoted as IO-10) and size tolerance of 2.5 nm in chloroform solution were purchased from Ocean Nanotech. Original concentration of maghimite was 21 mg/ml (0.43\% vol.).  MNPs were stabilized by oleic acid (C$_{18}$H$_{33}$COOH) monolayer shell of thickness $\approx 2$ nm to prevent coagulation. 

\section{Experiment}

\begin{figure}
\includegraphics[width=8.5cm]{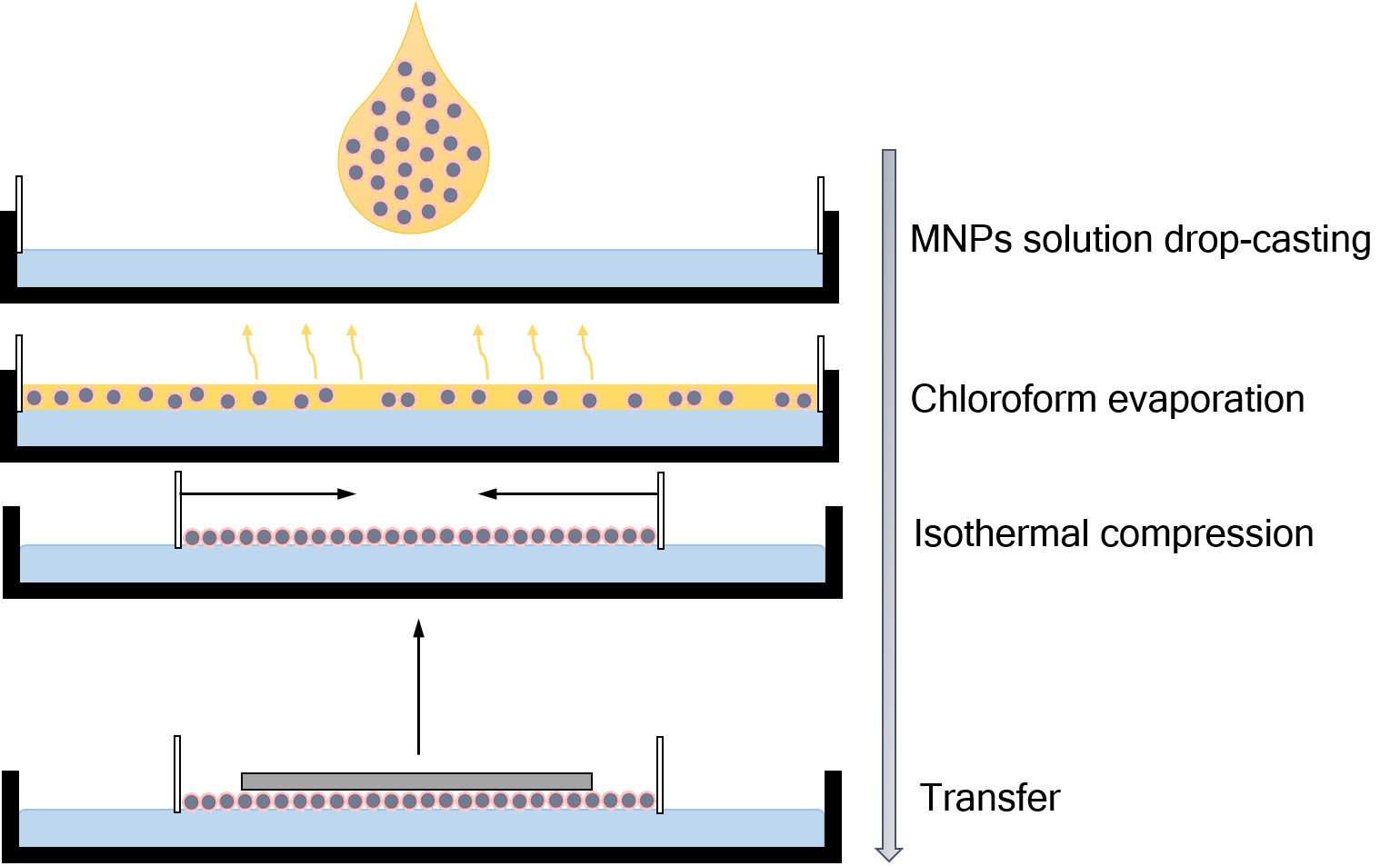}
\caption{Sketch of the Langmuir-Schaefer technique.}
\label{Fig1}
\end{figure}

The Langmuir MNP film was prepared in a custom-designed Langmuir trough, installed directly on the goniometer with the use of an active anti-vibration device Halcyonics MOD2-S. Maximum working area of the trough with fully opened barriers is 745 cm$^2$. The H$_2$O/IO-10 sample was prepared at room temperature by a micro syringe drop casting on different parts of the  water surface in the Langmuir trough. Prior to deposition, the as-purchased sample IO-10 was diluted to concentration 1.28 mg/ml by adding pure chloroform. No sonification was applied. More diluted solution allows for more homogeneous covering of the water surface and facilitates manipulations with the micro syringe. In total 0.3 ml of diluted sample was spread in approximately 5 min (3 micro syringes of 0.1 ml each). After the solvent evaporation the trough was sealed and filled with humid helium to minimize scattering on air and to compensate for evaporation of water. After the solvent evaporation, the film was compressed to reach a minimal area (61$\times$170 mm$^2$) by moving one barrier. Assembly of the LB film was followed by the LS deposition technique, as shown in Fig. \ref{Fig1}. 

XRR and GISAXS measurements were carried out at ID10 beamline of European Synchrotron Radiation Facility (ESRF, Grenoble, France), which is especially designed for studies on liquid surfaces \cite{smilgies2005troika}. Details of a basic principles and specific experimental approach of these surface-sensitive techniques can be found elsewhere \cite{renaud2009probing}. In the present experiments photons with wavelength $\lambda=1.54$ \AA~were used. XRR data were acquired by a one-dimensional position-sensitive detector (PSD) Vantec and two-dimensional PSD MARCCD 133 (2048 $\times$ 2048 pixels) was used for GISAXS measurement.

As a solid substrate we used gold covered Si crystal. A buffer titanium layer was introduced to improve adhesion of the gold layer. Top surface of gold was covered with a layer of organic molecules (1-pentadecanethiol) making it hydrophobic for better transfer of the MNP Langmuir layer which is also hydrophobic. Nominal (i.e. estimated from the deposition process) thicknesses of the gold and titanium layers were 10 nm and 5 nm respectively. In the following the substrate with deposited IO-10 layer is called sample Si/Ti/Au/IO-10.

It is worth noting that such multilayered substrate was chosen in view of our future plans to compare structure of Langmuir-Schaefer layers deposited on non-magnetic substrates and on magnetic substrates. In the later case the substrates will be similar to the present Si/Ti/Au but will contain an additional layer of cobalt between the gold and titanium layers. There gold will play a roll of capping material preventing cobalt oxidation. To separate reliably an effect of magnetic layer on the MNPs ordering it was decided to keep all other layers identical for both magnetic and non-magnetic substrates.

\section{Results and discussion}

\begin{figure}
\includegraphics[width=8.5cm]{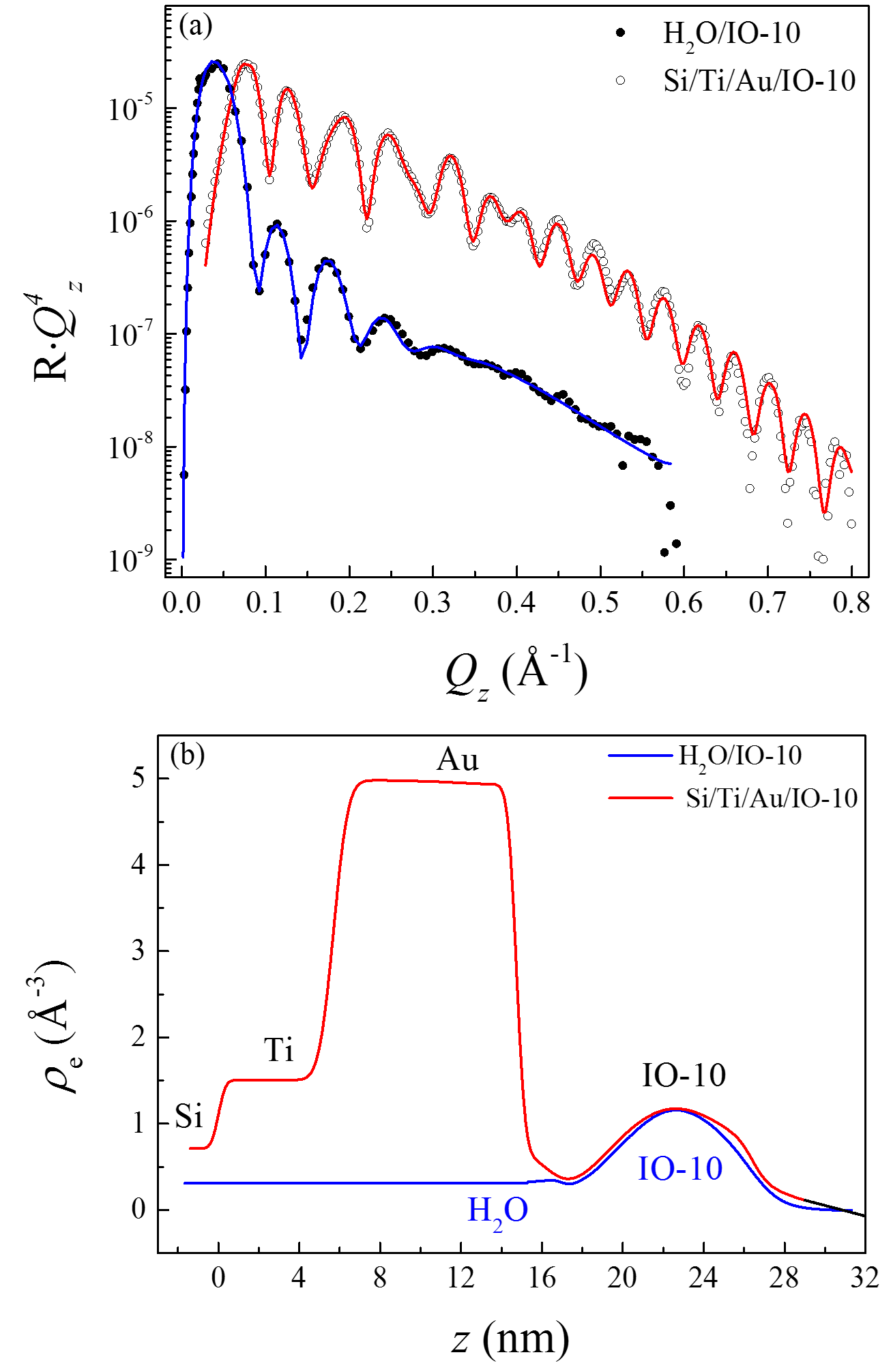}
\caption{(a) XRR data (symbols) and fitted (line) curves measured for the sample H$_2$O/IO-10 (filled circles) and sample Si/Ti/Au/IO-10 (open circles). The data for Si/Ti/Au/IO-10 are multiplied by 100 for clarity. (b) Electron density profiles of the sample H$_2$O/IO-10 (blue line) and Si/Ti/Au/IO-10 (red line).}
\label{Fig4}
\end{figure}

The transverse structure of the film was examined by XRR method which provides the information on electron density ($\rho_e$) of the film, as a function of a distance $z$ from an interface with a substrate. Distribution $\rho_e(z)$ is obtained by fitting the model XRR curves with GenX reflectivity tool \cite{bjorck2007genx}. Experimental XRR data (Fig.\ref{Fig4}) are represented as $RQ_z^4$ to emphasize visibility of the experimental data and the fit on all measured $Q_z$ range.

Presence of the IO-10 particles is manifested by a local increase of the electron density $\rho_e$ in the region $17.5 <z<28.5$ nm  (Fig.\ref{Fig4}b). Parabolic behavior of the electron density distribution in this region delivered by the fitting routine corresponds perfectly to an expected shape $\rho_{e}(z)$ for the case of a monolayer of spherical particles. Total thickness of IO-10 layer $d$ for both samples is 11.7 nm, what is corresponds to the nominal diameter of the IO-10 nanoparticles with collapsed surfactant shells. The maximum value of the electron density $\rho_e = 1.17$ \AA$^{-3}$ in the center of the layer was obtained also for both samples H$_2$O/IO-10 and Si/Ti/Au/IO-10 samples. Thus one can conclude that the transfer has not caused any additional defects to the monolayer, such as stacking of the particles into bi- or multi-layers.

The in-plane correlations between the nanoparticles in the IO-10 film were probed by GISAXS method. A 2D GISAXS pattern for the sample H$_2$O/IO-10 and cuts of the intensity distribution taken at grazing incidence and scattered angles $\alpha_i=\alpha_f=0.16^\circ$ are shown in Fig. \ref{Fig6}a,b.

It is well-known that monolayers of iron oxide MNPs tend to form a two-dimensional hexagonal close-packed (hcp) superlattice \cite{siffalovic2007self,siffalovic2008real,mishra2012self,guo2003patterned,lee2007vast, pauly2009large, wen2011ultra}. 

\begin{figure}
\includegraphics[width=8.5cm]{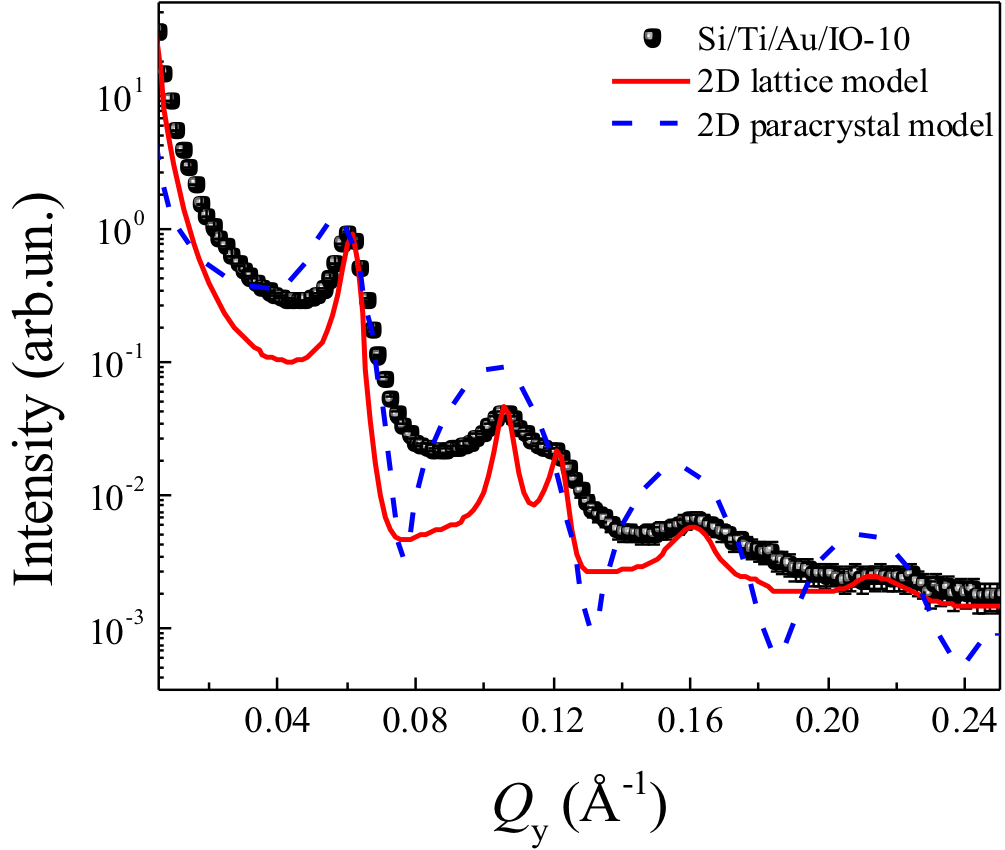}
\caption{Cut of the intensity of GISAXS experimental data from the Si/Ti/Au/IO-10 sample (symbols) and simulation with 2D hexagonal lattice model (red line) and 2D paracrystal model (dashed blue line). }
\label{Fig5}
\end{figure}

A corresponding set of Bragg peaks positions in a diffraction pattern can be described by the following relation:

\begin{equation}
Q^{hk}_y = \frac{2\pi}{d_{hk}},
d_{hk} = \frac{a}{\sqrt{\frac{4}{3}(h^2+hk+k^2)}},
\label{hk}
\end{equation}
where $Q^{hk}_y$ -- projection of the reciprocal space vector to the in-plane axis $y$, $h$ and $k$ are the Miller indices, $d_{hk}$ -- interplanar distance, $a$ is lattice constant.

One of the approximations of the nanoparticle assemblies is a paracrystal model, which exhibits a correlation between neighboring particles that  vanishes for the higher orders of the correlation function \cite{vegso2012nonequilibrium}. This approach is effectively used to describe ensembles with the short-range ordering. On the other hand, highly-ordered two-dimensional system can be described by the conventional diffraction theory based on the Bragg's law and Scherrer equation, which can provide lattice constant and mean crystal domain size \cite{heitsch2010gisaxs, smilgies2009scherrer}. We employed BornAgain software package \cite{durniak2015software} to simulate GISAXS signal using both 2D paracrystal and 2D lattice approximations. Because of the presence of the substrate layer the simulation was run using the Distorted wave Born approximation (DWBA). Cut of the measured GISAXS intensity from Si/Ti/Au/IO-10 sample and simulations are shown in Fig. \ref{Fig5}. We have found that paracrystal model (dashed blue curve in Fig.\ref{Fig5}) with a damping length equal to 0 and coherent domain size of 50 -- 1000 nm didn't fit the experimental results due to the rapid vanishing of the peaks coming from the hexagonal lattice leaving the form-factor oscillations alone at the $Q_y>0.07~\AA$. In second simulation MNPs were distributed along a 2D hexagonal lattice with a lattice constant $a$ shown in Table \ref{Tab} and average domain size $D$ determined from the experimental data using the Scherrer equation:

\begin{equation}
D = \frac{2\pi K}{\Delta Q^{hk}_y},
\label{sherrer}
\end{equation}

where $\Delta Q^{hk}_y$ is the full width at half-maximum (fwhm) of the Bragg peak and $K = 1.05$ is the Scherrer constant for two-dimensional hcp lattice \cite{lele1966influence}. For estimation of $D$ we used fwhm $\Delta Q$ of the most pronounced peak (10). Simulated scattering from domains with different crystallographic orientations was integrated in the simulation providing the whole set of Bragg peaks (solid red curve in Fig.\ref{Fig5}). Thus we can conclude that the sample Si/Ti/Au/IO-10 can be described as a two-dimensional polycrystal rather than a paracrystal model. Similar conclusion can be obtained for the H$_2$O/IO-10 sample. Polycrystallinity of the monolayer is naturally coming from the self-assembly process, which is passes through the step of connection of the individual ordered nanoparticle clusters to continuous layer \cite{vorobiev2015substantial}.

\begin{figure}
\includegraphics[width=8.5cm]{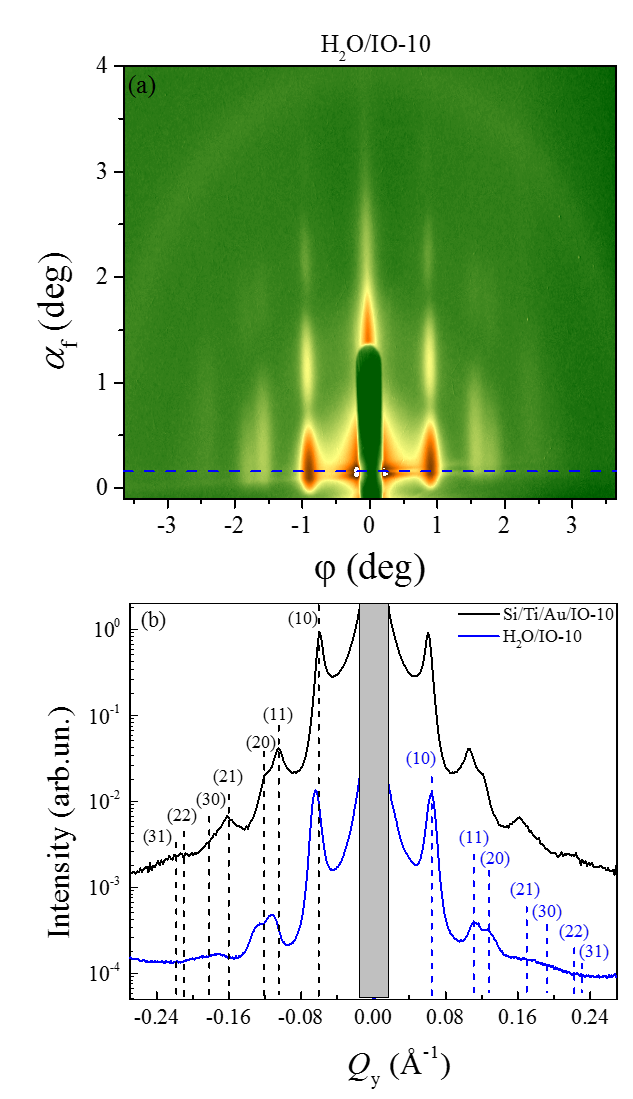}
\caption{(a) 2D GISAXS pattern for the sample H$_2$O/IO-10. Dashed blue line corresponds to $\alpha_i=\alpha_f=0.16^\circ$. (b) Cut of the GISAXS intensity from the samples H$_2$O/IO-10 (blue line) and Si/Ti/Au/IO-10 (black line) taken at $\alpha_i=\alpha_f=0.16^\circ$. The dashed lines correspond to the calculated peak positions.}
\label{Fig6}
\end{figure}

GISAXS data for the solid sample Si/Ti/Au/IO-10 is characterized by a higher signal to background ratio than the data for the sample on the liquid subphase: a set of seven Bragg peaks can be distinguished, while only five peaks are clearly observed in case of H$_2$O/IO-10 (Fig. \ref{Fig6}b). From the positions of the peaks $Q_{y}^{hk}$ a lattice constant $a_{1,2}$ was obtained for each sample by minimizing a value of squared deviation $\sigma=(d_{hk}^{exp} - d_{hk}^{calc})^2$, where $d_{hk}$ was calculated according to Eq.\ref{hk}. Small value of $\sigma$ confirms the hcp arrangement of the MNPs. All corresponding values are collected in Table \ref{Tab}. It was found that the lattice constant $a_2 = 11.94 \pm 0.10$ nm of the sample on the substrate is 6\% larger than $a_1 = 11.26 \pm 0.11$ nm in case of the sample on water. However, both in-plane lattice constants  are very close the thickness of the monolayer $d_1 = d_2 = 11.7$ nm obtained from the XRR data.

The average domain size of the two-dimensional superlattice is also found to be larger after the transfer ($D_2 = 95 \pm 5$ nm) comparing to the domain size on water ($D_1 = 71 \pm 7$ nm). 

\begin{figure}
\includegraphics[width=8.5cm]{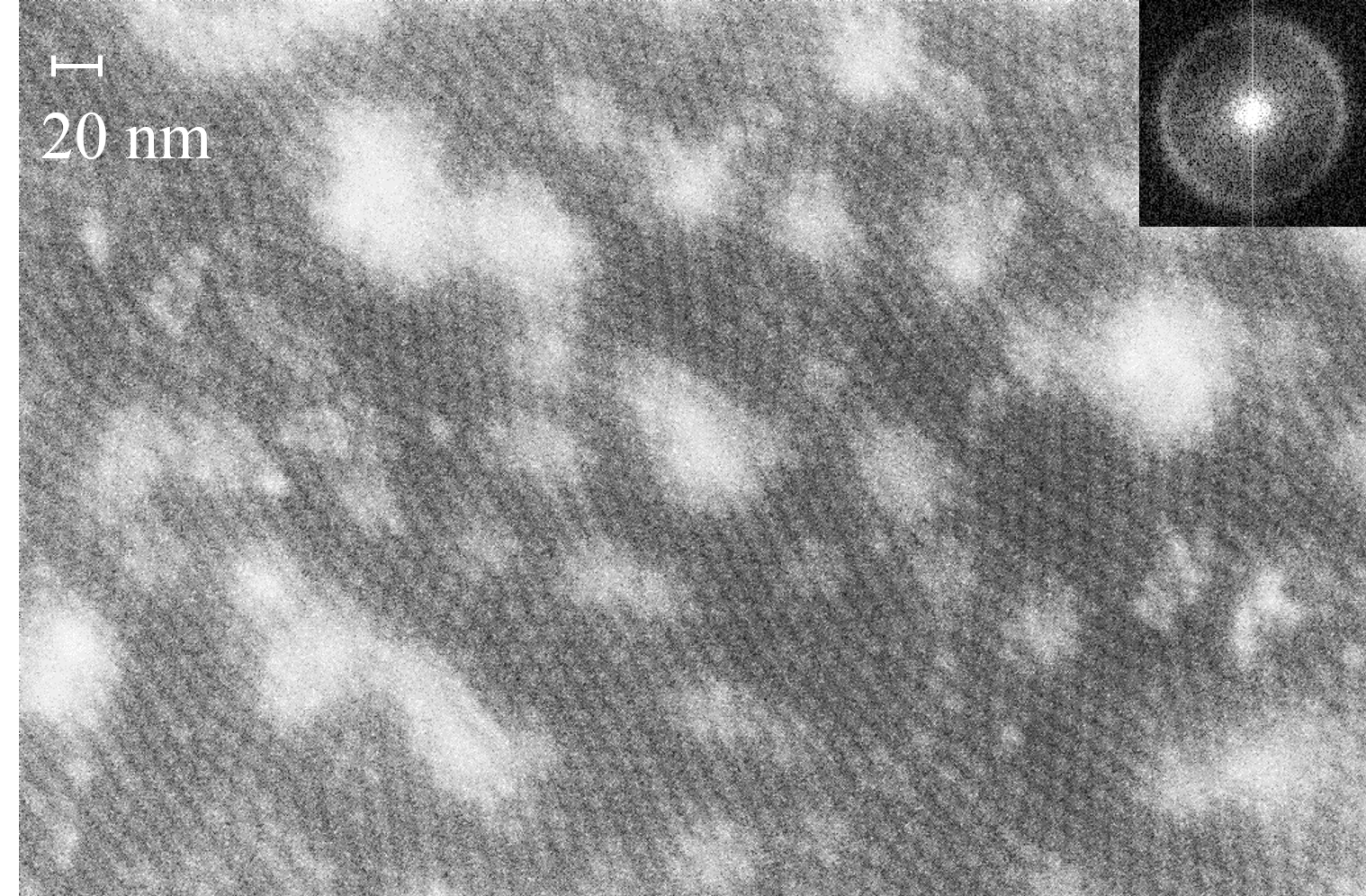}
\caption{SEM image of IO-10 nm particles deposited on a solid substrate. }
\label{Fig8}
\end{figure}

X-ray scattering data for Si/Ti/Au/IO-10 sample is supported by SEM measurements (Fig. \ref{Fig8}). Bright areas on the image are probably the regions contaminated by free surfactant molecules. From a diameter of a ring on Fourier transform of raw SEM image (shown in inset of Fig. \ref{Fig8}) one can find a value $d_{hk}=10.8 \pm 1.3$ nm corresponding to the Bragg peak (10) what is in good agreement with the corresponding value obtained by GISAXS.

Summarizing all presented results one can conclude that a short-range ordering, which is characterized by the interparticle distance, is better in the layer on the water surface. In contrast, the average domain size is slightly bigger in the layer on the solid substrate. However, both systems can be described by the two-dimensional polycrystal model rather than paracrystal. The slight difference in ordering of the sample on the liquid subphase can be explained by a perturbing influence of capillary waves always present on free water surface. Apparently they prevent formation of large monocrystal domains but, on the other hand, they stimulate local ordering of the particles. After the transfer the lattice constant increases subtly what may indicate that a balance of forces acting between the particles shifts slightly in favor of repulsive forces.

\begin{table*}
\begin{center}
\begin{tabular}{llllll}
\hline
Sample &($h$ $k$)& $d^{calc}_{hk}$, nm& $d^{exp}_{hk}$, nm & $a$, nm & $D$, nm\\
\hline
H$_2$O/IO-10& (10) & 9.83 & 9.83 &$11.26 \pm 0.10$ & $71 \pm 5$\\
 & (11) & 5.68 & 5.61\\
 & (20) & 4.92 & 4.91\\
 & (21) & 3.72 & 3.64\\
 & (30) & 3.28& 3.26\\
\hline
Si/Ti/Au/IO-10& (10) & 10.38 & 10.38 &$11.94 \pm 0.11$ & $95 \pm 7$\\
 & (11) & 5.99& 5.94\\
 & (20) & 5.19& 5.17\\
 & (21) & 3.92& 3.86\\
 & (30) & 3.46& 3.49\\
 & (22) & 2.99& --\\
 & (31) & 2.88& --\\
\hline
\end{tabular}
\end{center}
\caption{Measured ($d^{exp}_{hk}$) and calculated ($d^{calc}_{hk}$) interplanar distances for the samples H$_2$O/IO-10 and Si/Ti/Au/IO-10. The average domain size was determined according to Eq.\ref{sherrer}. Peaks (22) and (31) for the sample Si/Ti/Au/IO-10 are visible at GISAXS cut in Fig. \ref{Fig6} but not resolved.}
\label{Tab}
\end{table*}

\section{Conclusion}

Using the Langmuir-Blodgett method a monolayer of magnetic highly-monodisperse iron oxide nanoparticles of 10 nm was assembled. Both in-plane and out-of-plane structure of the monolayer was controlled by X-ray reflectometry and grazing-incidence small-angle scattering. It was shown that the nanoparticles form a two-dimensional hexagonal close-packed superlattice. The same scattering methods were used to study Langmuir-Schaefer monolayer deposited on solid substrate Si/Ti/Au. Resulted monolayer on the solid substrate was found to be the same or even better quality as the original sample on water subphase confirming a good adhension of the nanoparticles by gold-thiol mediated layer. However, the in-plane lattice constant of the two-dimensional superstructure of the deposited sample is 5\% larger than lattice constant of the superlattice on water subphase. 

\subsection*{Acknowledgements} Authors thank European Synchrotron Radiation Facility for the provided beamtime and technical assistance. This work was supported by the Russian Foundation for Basic Research (grant 14-22-01113-ofi m).

\bibliographystyle{model1a-num-names}
\bibliography{sample.bib}

\end{document}